\documentclass[sigconf]{acmart}
\usepackage{subcaption}
\usepackage{booktabs}
\usepackage{float}
\usepackage[draft,inline,nomargin,index]{fixme}
\usepackage{graphicx}
\usepackage{caption}
\usepackage{hyperref}
\hypersetup{
    colorlinks=true,
    linkcolor=red,
    filecolor=magenta,      
    urlcolor=cyan
    }
\usepackage{soul}

\fxsetup{theme=colorsig,mode=multiuser,inlineface=\itshape,envface=\itshape}

\AtBeginDocument{%
  }

\copyrightyear{2026}
\acmYear{2026}
\setcopyright{cc}
\setcctype{by}
\acmConference[CHI '26]{Proceedings of the 2026 CHI Conference on Human Factors in Computing Systems}{April 13--17, 2026}{Barcelona, Spain}
\acmBooktitle{Proceedings of the 2026 CHI Conference on Human Factors in Computing Systems (CHI '26), April 13--17, 2026, Barcelona, Spain}
\acmPrice{}
\acmDOI{10.1145/3772318.3790489}
\acmISBN{979-8-4007-2278-3/2026/04}

\begin{document}

\title{Amplifying Rural Educators' Perspectives: A Qualitative Study on the Impacts of Generative AI in Rural U.S. High Schools}

\author{Shira Michel}
\email{michel.sh@northeastern.edu}
\orcid{0009-0003-2143-991X}
\affiliation{%
  \institution{The Roux Institute at Northeastern University}
  \city{Portland}
  \state{Maine}
  \country{USA}
}

\author{Benjamin Taylor}
\email{ben.taylor@katabasis.org}
\orcid{0009-0003-1576-6818}
\affiliation{%
  \institution{Katabasis}
  \city{Garner}
  \state{North Carolina}
  \country{USA}
}

\author{Sabrina Parra Díaz}
\email{sabrina.parra@katabasis.org}
\orcid{0009-0001-9399-9444}
\affiliation{%
  \institution{Katabasis}
  \city{Garner}
  \state{North Carolina}
  \country{USA}
}

\author{Joseph B. Wiggins}
\email{joseph.wiggins@katabasis.org}
\orcid{0000-0002-6170-7080}
\affiliation{%
  \institution{Katabasis}
  \city{Garner}
  \state{North Carolina}
  \country{USA}
}

\author{Ed Finn}
\email{edfinn@asu.edu}
\orcid{0000-0002-6282-9995}
\affiliation{%
  \institution{Arizona State University}
  \city{Tempe}
  \state{Arizona}
  \country{USA}
}

\author{Mahsan Nourani}
\email{m.nourani@northeastern.edu}
\orcid{0000-0002-8823-9635}
\affiliation{%
  \institution{The Roux Institute at Northeastern University}
  \city{Portland}
  \state{Maine}
  \country{USA}
}

\renewcommand{\shortauthors}{Michel et al.}

\renewcommand{\shorttitle}{A Qualitative Study on the Impacts of Generative AI in Rural U.S. High Schools}

\begin{abstract}
Recent breakthroughs in Generative AI (GenAI) are reshaping educational landscapes, presenting challenges and opportunities. While all contexts present unique challenges, rural schools are historically under-resourced, facing persistent technology-related barriers. To understand and reduce these barriers, we studied 31 rural high school educators across three U.S. states to examine their use of GenAI and understand how GenAI introduces new challenges, opportunities, and may exacerbate existing educational barriers. Results show while rural educators use GenAI to streamline teaching tasks, existing resource disparities restrict meaningful integration. Through rural educators' voices, we reveal issues like infrastructure barriers, resistance to adoption, and lack of AI literacy training create significant obstacles. Nonetheless, educators envision GenAI can support themselves and their students, but findings emphasize the need for rural-specific design approaches. As a community, embracing inclusive GenAI design and re-examining assumptions about technology adoption in under-served educational contexts is essential to reducing barriers rather than widening them.
\newline
\textbf{Supplemental Material} is open-sourced and available at \\
\href{https://osf.io/8hckv/?view_only=1830d5f10bda49df81da60a42df10225}{https://osf.io/8hckv/}.
\end{abstract}



\maketitle

\section{Introduction}
\label{sec:intro}

Few technologies have arrived with the immediacy and transformative power of Generative AI, representing one of the most rapid and far-reaching technological disruptions, and reshaping fundamental assumptions about Human-Computer Interaction (HCI) in a matter of years rather than decades.
K-12 educational institutions are among those most significantly affected~\cite{han2024teachers, velander2024artificial}.
Students are navigating a formative period of skill development and academic growth~\cite{demir2022examination}, making the ways they engage with GenAI especially consequential.
GenAI's appealing promise of efficiency can interrupt the development of essential skills, endurance, and self-efficacy, creating what Cecilio-Fernandes and Sanders~\cite{cecilio2025hallucination} call a ``\textit{hallucination of learning}''---a false sense of understanding without genuine skill acquisition. 
Educators are experiencing unprecedented uncertainty and anxiety as GenAI fundamentally disrupts their pedagogical worldviews~\cite{ahn2025exploring, harvey2025don, tan2024more}.
Recent studies reveal AI-based chatbots may decrease reasoning abilities and cognitive skills~\cite{kosmyna2025your} while promoting ``\textit{metacognitive laziness}''~\cite{fan2025beware}, adding on to existing concerns about plagiarism and decreasing self-efficacy.
Unlike previous technological advancements (e.g. calculators, social media) contained within specific contexts, GenAI has become embedded in everyday tools---from Google search~\cite{reid2024genai} to word processors ~\cite{ms2025welcome, google2025collab}---creating new challenges for educators to determine its classroom influence.
As a result, educators have to navigate new pedagogical terrain.

The emerging picture is not uniformly dystopian.
Recent studies showcase opportunities and creative interventions that harness GenAI's potential while minimizing its risk, such as equipping students and teachers with AI literacy skills~\cite{cao2025ai, dangol2025ai, han2025empowering} to recognize errors produced by GenAI systems~\cite{dangol2025ai}, suggesting thoughtful implementation can reimagine educational possibilities.
Yet this optimistic research landscape implicitly assumes schools have adequate resources, reliable technical infrastructure, and institutional support.
As a result, rural educational environments, which operate under fundamentally different constraints, including limited connectivity, chronic underfunding, and staffing challenges, face distinct barriers to technology integration~\cite{rude2018policy, kolbe2021additional}.
Educational research has tended to prioritize urban and suburban networks~\cite{penn2024restore, aasa2023reimagine, pollock2024leveraging, tomanek2005points}  and rarely specifies the geographic characteristics of study populations~\cite{belghith2024testing}, leaving less visibility on insights from how rural educational communities and stakeholders experience, adapt to, and envision using these emerging technologies.
This gap matters---nearly one-fifth of K-12 American public school students attend rural schools (around $\sim10$ million total students)\footnote{Per data from \textit{National Center for Education Statistics: \url{https://nces.ed.gov/programs/coe/indicator/lcb/school-choice-rural}}, and Issue by \textit{National Education Association: \url{https://www.nea.org/advocating-for-change/action-center/our-issues/rural-schools}, both accessed on: September 9, 2025}.}, making it essential to include their educators' voices in conversations about GenAI's educational future.

This work was therefore motivated by a desire to amplify rural educators' voices and perspectives from a balanced stance---neither advocating for nor opposing GenAI adoption, but foregrounding their \textit{authentic} experiences and viewpoints. 
Our investigation was guided by the following research questions:
 \begin{quote}
    \textbf{RQ1}: How do rural contexts shape high school educators’ experiences with GenAI integration?\\
    \textbf{RQ2}: How are rural high school educators navigating GenAI integration in their schools and classrooms?\\
    \textbf{RQ3}: What do rural high school educators envision GenAI could support them with?
 \end{quote}

To address these questions and gaps, we designed a qualitative user study, including an online survey followed by in-depth interviews, with rural high school educators across three states: Arizona, Maine, and North Carolina.
We analyzed the data using a grounded theory-informed approach with open coding~\cite{khandkar2009open} to identify emerging themes.
We then conducted a secondary analysis through a critical rural theory lens~\cite{thomas2013critical}, examining connections between educators' responses and theoretical concepts of power relations and systemic inequities.
To the best of our knowledge, this is among the first U.S. studies examining rural educators' GenAI usage, perspectives, and integration barriers in high school contexts.
We focus on high school education because students at this level have developed the critical thinking skills necessary for meaningful technology engagement ~\cite{demir2022examination}, and their educators play crucial roles in college and career preparation ~\cite{johnston2021believe}.
Furthermore, research shows technology-supported learning enhances high school student outcomes and creates supportive learning environments ~\cite{fikriyah2022use}, making these settings important for understanding GenAI's educational implications.


Based on our findings, rural educators' engagements with GenAI reveal a landscape of structural contraints and resourcefulness.
In spite of persistent barriers---such as limited infrastructure, high workloads, and technological isolation---that affect access, these educators bring a unique perspective to GenAI integration, informed by years of creative problem-solving in their rural contexts.
Their responses ranged from cautious considerations about GenAI's immediate relevance to appreciation for the ways their technological isolation provides a temporary buffer from needing to engage with GenAI in their schools. 
Most strikingly, they envisioned transformative applications when given the opportunity to speculate: adaptive learning systems for multi-grade classrooms, AI-powered curricula incorporateing local knowledge, and digital tools addressing chronic staffing shortages through personalized instructions.
These reveal a fundamental inequity and power imbalance in educational technology development---rural educators possess the adaptive expertise and contextual creativity essential for thoughtful GenAI implementation, yet their insights remain underrepresented in design processes and policy decisions.
Amplifying rural voices not only benefits these communities, but also unlocks untapped potential that can inform innovation and decisions around GenAI applications.

\textbf{Centering rural educators' perspectives not only illuminates broader GenAI opportunities and challenges across \textit{all educational settings} but also the \textit{unique ways these technologies intersect with rural community contexts}.}
This research represents a crucial step t addressing inequities in educational technologies and creating truly inclusive innovation pathways with rural stakeholders in mind.
To support this transformation, our complete methodological toolkit---including our codebook, interview protocol, and survey instruments---is open-sourced and available to empower future researchers.


\section{Defining Rural}
To situate this work within the rural education research space, it is important to first define what we mean by ``rural''.
Unfortunately, given the size and diversity of the US, no single definition of rural can adequately capture the varied contexts across different states, including the federal classifications which have these same limitations.
The 2020 Census Bureau~\cite{us2024rural} classifies\footnote{The U.S. Census Bureau does not actually define ``rural'' as stated in their most recent survey~\cite{us2020understanding}.} ``rural'' as ``\textit{any population, housing, or territory \textbf{not} in an urban area}''.
A territory qualifies as an ``urban area'' if it encompasses at least 2,000 housing units or has a population of at least 5,000\footnote{This is a new definition different from the 2010 U.S. Census Bureau's definition consisting both urbanized areas, a population of at least 50,000, and urban clusters, a population of at least 2,500 and less than 50,000 ~\cite{us2020understanding, define_rural}. Rural classifications were further divided into fringe, rural, distant rural, and remote rural. Fringe rural is less than 5 miles from an urbanized area and less than 2.5 miles from an urban cluster;
distant rural is between 5 and 25 miles from an urbanized area and  between 2.5 and 10 miles from an urban cluster;
remote rural is more than 25 miles from an urbanized area and more than 10 miles from an urban cluster ~\cite{nces2022locale}. 
Although the Census Bureau no longer uses these definitions and classifications, other organizations use these criteria, such as the National Center for Education Statistics (NCES)~\cite{nces2022locale}. This inconsistency shows how ``rural'' is ill defined and what counts as ``rural'' varies greatly. } ~\cite{us2024rural}.

Throughout this paper, we will describe our rural spaces in regards to both their housing units and population density to help capture and define the rural context. 
Most importantly, however, we center participants' own definitions of ``rural'' based on their lived experiences.

Another important component of understanding rurality is through understanding \textit{rural consciousness} ~\cite{walsh2012putting}, a theoretical shared identity of rurality.
Rural consciousness encompasses (1) a sense of belonging to a rural geographical location, (2) embracing rural values and lifestyles, and (3) espousing a belief of distributive social injustice where those in power often ignore rural concerns while prioritizing urban and suburban populations.
This third element of rural consciousness connects directly to \textit{critical rural theory}~\cite{thomas2013critical}, which situates rural studies within a critical framework that examines mechanisms of cultural and structural domination. 
Key concepts of critical rural theory include \textit{urbannormativity}, the perception of urban life as ``normal'' and rural life as ``abnormal''~\cite{thomas2013critical}, and \textit{place-structuration}, where culturally driven meaning-making marginalizes rural spaces and identities~\cite{thomas2013critical}. 
Within educational settings, critical rural theory also showcases how integrative institutions, such as schools and nonprofits, can reproduce urban-centric ideologies~ \cite{seale2013studies} and how standardized educational policies, including school consolidation and state-mandated curricula, can distance rural students from their local contexts~\cite{cervone2017reproduction}.
These critical lenses, combined with educators' own community definitions, ground our results and discussions in a human-centered and data-driven understanding of rurality.

\section{Related Work}
\label{sec:related}
In this section, we provide context for positioning GenAI as a socio-technical artifact with insights from integrating earlier educational technologies and from critical theories. 
We review current research in the HCI community on how GenAI is being used by both students and educators as well as the educational settings in which these studies take place.

\subsection{GenAI: a Socio-Technical Artifact}
GenAI is a socio-technical artifact whose design, development, and use  shaped by both its technical affordances and contexts in which it operates~\cite{roe2025generative,seaver2018should,ehsan2020human}. 
In other words, the meanings and consequences of GenAI cannot be understood apart from the values and power relationships it embodies~\cite{mitra2024sociotechnical}.
GenAI reflects the priorities of dominant social groups and institutional structures, influencing how knowledge is produced, applied, and valued~\cite{feenberg2008critical}. 
When only the narrow interests of these groups are considered, those excluded from participation often bear the burden of unintended or undesirable consequences.

The adoption of GenAI in education mirrors the pathways established by earlier educational technologies, such as standardized testing, learning management systems, and the internet~\cite{assessment-equity, onlinePlatforms-equity, digitalEd-equity}.
These technologies historically reorganized schooling around performance metrics rather than educational equity~\cite{onlinePlatforms-equity, coates2005critical, national1992lessons, edTech-sociocultural}.
For example, the ties between test scores and funding have created systemic conditions in which resources are unevenly distributed and differences between school contexts are amplified~\cite{edDisparities}.
Similarly, the governance of digital learning infrastructures has often been mediated by assumptions of broadband, obscuring the realities of under-resourced communities~\cite{antoninis2023global}, such as those in rural areas.

Additionally, as an embodied technology, GenAI's value in practice depends on whether educators and students find it useful, trustworthy, and culturally relevant ~\cite{viberg2024explains,cukurova2023adoption,gardan2025adopting, nazaretsky2025critical}. 
Prior work suggests AI's effectiveness in education depends on its situated use, where meaning emerges through human-AI collaboration between educators and students~\cite{sharples2023towards, lee2023collaborative, holstein2022designing}. 
Without thoughtful integration, technologies like GenAI are at risk of being resisted or producing alienating effects instead of meaningful learning outcomes.
Drawing on critical rural theory~\cite{thomas2013critical}, we investigate how GenAI functions as a socio-technical artifact within rural educational contexts, exploring the interplay between existing educational inequities and GenAI implementation---how these inequities influence GenAI adoption and outcomes, and how GenAI, in turn, may either reproduce or transform these inequities while opening new avenues for inclusive participation.

\subsection{GenAI Applications and Studies in K-12}
Recent research has shown potential GenAI applications in formal and informal educational settings.
For students, GenAI could serve as virtual teaching assistants that support~\cite{lieb2024student, kazemitabaar2024codeaid, taneja2024jill} or have the role of the instructor~\cite{van2023help, yalccin2022intelligent, cai2024advancing, steenstra2024engaging}, and even trainees, where students reinforce learning by teaching AI ~\cite{shahriar2023and, jin2024teach}. 
These agents could support students in ideation~\cite{goldi2024intelligent, lawton2023drawing} and explaining concepts~\cite{chen2024bidtrainer} via personalized learning materials, including exercises (e.g.,~\cite{ooge2023steering, ooge2022explaining}) and interactive activities (e.g.,~\cite{karaosmanoglu2024language}).
For educators, GenAI can potentially help design quizzes and homework~\cite{elkins2024teachers, wang2022towards}, develop course materials ~\cite{abolnejadian2024leveraging, han2024teachers}, automate administrative tasks~\cite{oh2024exploring}, and assist with classroom management~\cite{yang2023pair}.
GenAI offers great potential for student engagement and learning across disciplines. 
Recent work demonstrates numerous examples of GenAI providing coding assistance~\cite{chen2024chatscratch}, enhancing STEM learning through storytelling and problem-solving~\cite{cheng2024scientific, zhang2024mathemyths}, and supporting language learning and social studies education~\cite{karaosmanoglu2024language, hedderich2024piece}.
Moreover, GenAI has been used in interdisciplinary contexts ~\cite{fan2025litlinker} to bridge subject domains and literacies.
Beyond the academic content, GenAI can help foster socio-emotional growth and empathy through behavioral learning activities ~\cite{lo2025noel}.

To understand current GenAI roles and potentials within K-12 contexts, several recent studies have focused on surveying and interviewing educators~\cite{belghith2024testing, van2023help, xiao2024exploring, han2024teachers, tan2024more, oh2024exploring, ahn2025exploring, hartman2023rural, harvey2025don}.
For example, Han et al. ~\cite{han2024teachers} investigated how elementary school stakeholders perceive GenAI’s role in supporting writing instruction and explored ways GenAI could be designed to strengthen literacy learning. 
Tan et al. ~\cite{tan2024more} interviewed middle and high school teachers to identify what information and resources they need to meaningfully integrate GenAI into their classrooms.
These studies are collectively improving researchers' understanding of the current attitudes and perceptions toward GenAI.
Across these studies, many educators acknowledge GenAI's potential to enhance both their teaching and students' learning experiences, while being concerned about aligning GenAI with pedagogical goals, its practical limitations, and their own confidence in using the technology effectively~\cite{tan2024more, ahn2025exploring, harvey2025don}.
Ahn and Lim ~\cite{ahn2025exploring} uncover how K-12 physical education teachers were concerned about data security and privacy when students or themselves use GenAI tools, and noted the challenges of accommodating educators with varying levels of technology proficiency in identifying GenAI's limitations.
Similarly, Harvey et al.'s ~\cite{harvey2025don} interviews with K-12 educators revealed their concerns about the broader harms of interacting with these systems, especially when they felt unable to personally mitigate those risks.
In response to these challenges and concerns, more recent studies emphasize the importance of equipping both students and teachers with AI literacy skills ~\cite{cao2025ai, han2025empowering, dangol2025ai} and the ability to critically identify and evaluate errors produced by GenAI systems ~\cite{dangol2025ai}. 
Other work ~\cite{belghith2024testing, han2024teachers, zhang2024mathemyths, he2025carlitos, pu2025can, ahn2025exploring, lo2025noel} highlights the value of participatory approaches that center the perspectives of educators and students, such as how GenAI can be made more culturally responsive ~\cite{he2025carlitos} and more relevant across different grade levels ~\cite{pu2025can, belghith2024testing}.

It is important to note much of the existing research on AI and GenAI in education focuses on urban-suburban schools or fails to differentiate the context. 
When rural schools are included, they are often represented by smaller sample sizes 
and analyzed in aggregate with other educational communities ~\cite{ravi2025co}.
This approach overlooks the distinct contexts of rural schools, making it harder to understand how GenAI might be meaningfully integrated.
From the perspective of critical rural theory ~\cite{thomas2013critical}, this lack of contextual distinction reinforces the pattern in which rural communities are treated as homogeneous with their urban and suburban counterparts. 
Not only do voices from these communities get lost in the aggregate, the development of technologies and policies are also limited to respond to rural needs.
As a result, rural perspectives remain underrepresented in current scholarship.
This gap is likely due to the geographic proximity of well-resourced research universities and institutions that frequently collaborate with urban-suburban schools ~\cite{pollock2024leveraging, tomanek2005points, aasa2023reimagine, penn2024restore}.
Furthermore, standards and policies governing GenAI use in educational settings are not well-established and vary across states ~\cite{fitzgerald2025majority, ncsl2025ai, dawson2025how}.
Such inconsistencies create challenges for all schools, but in rural schools, distinct infrastructural barriers such as broadband access~\cite{horrigan2015home, horrigan2010broadband} compound upon these challenges and therefore, there is a need for GenAI approaches designed with diverse educational contexts in mind.

\section{Methodology}
\label{sec:method}
We surveyed and interviewed rural high school educators to examine how GenAI has influenced their classrooms, teaching experiences, and student learning, as well as the opportunities they envision for its use.
This research was approved by our institutional review board (IRB).
\textbf{All the study instruments are open access and can be found in the supplemental material.}


\subsection{Online Survey}
\subsubsection{Design}
To understand the interplay among the social, infrastructural, and educational contexts of rural schools and GenAI, we designed an online survey, including multiple-choice and open-ended questions.
These questions were informed by prior work that suggests educators' willingness and ability to engage with new technologies depend on how well the tools align with their tasks, the support and resources available, and their familiarity and perceived usefulness (e.g., ~\cite{cheah2025stem, icf2009communities, han2024teachers}).

The survey included two sections on AI and GenAI, respectively.
Each section provided a brief definition adapted from IBM research~\cite{martineau2023genai}, followed by two multiple-choice questions: one asking participants to identify examples of the aforementioned technology to assess their understanding and familiarity, and another asking about their usage frequency and contexts.
Examples of applications for question one included \textit{``Smart replies in emails or texts''} for AI and \textit{``ChatGPT'', ``Midjourney'', ``DALL-E''}, for GenAI, adapting from prior works ~\cite{cai2024advancing, collie2024teachers, cheah2025stem}.
Participants who indicated any familiarity were asked to elaborate on the specific ways they used these technologies and how these recent advancements 
have impacted their classrooms and teaching.

The survey then included two multiple-choice questions framed around speculative GenAI usage scenarios within the context of rural schools.
These scenarios were informed by prior research on GenAI in non-rural educational contexts ~\cite{abolnejadian2024leveraging, han2024teachers, shahriar2023and, oh2024exploring, tan2024more, yang2023pair, taneja2024jill, belghith2024testing, ooge2022explaining, ooge2023steering} and were intended to gauge interest in possible future applications.
One of these use cases was ``\textit{GenAI could take on the role as a teaching assistant.}'' ~\cite{taneja2024jill}, and for students, ``\textit{GenAI could serve as a tutee and students could portray the role as the tutor.}''~\cite{shahriar2023and}.

To assess attitudes and perceptions towards GenAI, we adapted questions from the Computer Anxiety Rating Scale (CARS)~\cite{heinssen1987assessing} and Technology Acceptance Model (TAM)~\cite{davis1989perceived}, modifying them to focus specifically on GenAI and framing them within the context of rural education.
These validated questionnaires were chosen over specific AI questionnaires to allow for comparison with broader patterns of technology adoption in rural communities.
Both questionnaires captured core concepts in technology-related anxiety and perceived usefulness that are directly applicable to understand rural educator's potential engagement with GenAI.
The CARS and TAM questionnaires used a 5- and 7-point Likert scale, respectively, with optional text-fields to elaborate on their overall responses.

We included open-ended questions for educators to reflect on their daily experiences~\cite{icf2009communities, woodcock2022teacher, nwoko2023systematic, eryilmaz2025teacher}, 
including their favorite aspects of teaching, the resources they rely on (particularly technological), and the tedious tasks that divert attention from teaching and interacting with students. 
We asked them to assess their ability to support students with diverse needs given the current resources and support available in their schools ~\cite{turner2014inclusive}, on a 7-point Likert scale.
These questions helped us gain deeper insights into the realities of their day-to-day work.

The study concluded with demographic, geographic, and teaching background questions. 
Demographic questions captured participants' self-reported age, gender, race, and ethnicity. 
Questions on geography and rurality assessed participants' self-reported rurality of their school communities and their state location.
Teaching experience questions examined length of service, subject areas taught, and grade levels.

Two attention checks were randomly embedded within the main body of the survey to ensure data quality.
These questions were excluded from the analysis, and were only included to help remove inattentive participants who failed either of the questions.

\subsubsection{Procedure}
Participants completed the survey in a single, approximately 15 minute session via Qualtrics\footnote{\url{https://www.qualtrics.com/}} between August and September 2024.
After providing informed consent, participants filled out the main survey questions and post-survey, followed by demographic and teaching information.
Those who completed the survey could optionally enter a raffle to win one of six \$15 (USD) gift cards by providing their email addresses for contacting winners.
At the end of the study, educators could also provide their email address for future communications, including taking part in follow-up interviews.
\begin{figure*}
\begin{minipage}[t!]{.45\textwidth}
    \vspace{0.5 cm}
    \centering
    \includegraphics[width=.95\textwidth]{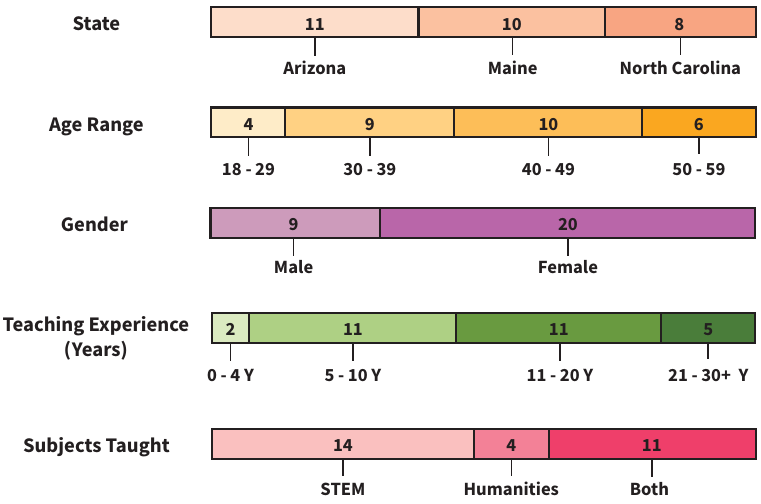}
\end{minipage}\hfill %
\begin{minipage}[t!]{.5\textwidth}        
    \centering
    \resizebox{\linewidth}{!}{%
        \begin{tabular}{@{}cccccc@{}}
        \toprule
        \multicolumn{6}{c}{\textbf{Interview Participants' Demographics}} \\ 
        \midrule
        \textbf{PID} & \textbf{State} & \textbf{Age Range} & \textbf{Gender} & \textbf{Teaching Experience} & \textbf{Subjects Taught} \\ 
        \midrule
        I1 & Arizona  & 40 - 49 & Female & 5 - 10 Y & STEM \\
        I2 & Arizona & 30 - 39 & Female & 0 - 4 Y & Both \\ 
        I3 & Maine & 50 - 59 & Female & 21 - 30 Y & STEM \\
        I4 & Maine & 40 - 49 & Male & 21 - 30 Y & Humanities \\ 
        I5 & North Carolina & Undisclosed & Undisclosed & 11 - 20 Y & Humanities \\
        I6 & North Carolina & Undisclosed & Undisclosed & 21 - 30 Y & STEM \\
        \bottomrule
        \end{tabular}%
    }
\end{minipage}
\par
\medskip
\noindent
\begin{minipage}[t!]{.49\textwidth}
  \centering
  \captionof{figure}{Participant demographics for the online survey ($N=29$). Rectangle widths and internal numbers indicate the count of participants who self-reported each demographic category.}
  \label{fig:survey-participants}
  \Description{A chart with five bars of the participants from the online survey. The bars are segmented in length by self-reported counts. Each bar represents each self-reported demographic category: state location, age range, gender, teacher experience range in years, and subjects taught.}
\end{minipage}%
\hfill
\begin{minipage}[t!]{.49\textwidth}
  \centering
  \captionof{figure}{Interview participants demographics ($N=6$). Two participants did not complete the survey, so only demographic information known for all interviewees is reported.}
  \label{tab:interview-participants}
  \Description{A table of the interview participants' demographics. This includes their participant identifier and self-reported state location, age range, gender, teaching experience range in years, and subjects taught.}

\end{minipage}
\end{figure*}

\subsubsection{Participants}
We recruited high school educators from rural communities in three states (Arizona, Maine, and North Carolina) that represented diverse rural contexts within the US.
Participants were required to be at least 18 years old, feel comfortable using an online interface to read and answer questions, speak English fluently, and be currently teaching in a rural high school. 
Educators from all subject areas were eligible to participate.
Consistent with recommendations by Hardy et al. ~\cite{hardy2019rural}, effective recruitment in rural communities involves partnering with local organizations and using existing social networks that respect local values and build trust, which can be difficult for external researchers and large institutions to establish.
Following this guidance, we distributed recruitment information through our networks using email and academic listservs across all three states and snowball sampling.
In addition, we partnered with the Maine Mathematics and Science Alliance (MMSA)\footnote{\url{https://mmsa.org/}} to reach rural educators in Maine.
Through these means, we recruited a total of 32 participants from Arizona, Maine, and North Carolina.
Three educators failed at least one of the attention checks.
From the remaining 29 participants, 11 educators were from Arizona, 10 educators from Maine, and 8 educators from North Carolina (see ~\autoref{fig:survey-participants} for demographic breakdown overview).

\subsection{Deepening Insights with Interviews}
Survey participants who expressed interest were invited to follow-up interviews for a more in-depth conversation about the impact of GenAI on their rural educational contexts.
Due to non-response bias~\cite{sheikh1981investigating} from participants who completed the survey, we additionally recruited 2 new participants via email, who did not originally complete the survey. 
With 4 returning participants, this led to a total of 6 interview participants (see~\autoref{tab:interview-participants} for their demographic breakdown).
All interview participants were compensated with a digital \$20 (USD) gift card.

\subsubsection{Interview Protocol}
\label{sec:interview}
We conducted semi-structured interviews over Zoom\footnote{\url{https://www.zoom.com/}} between October and November 2024, with each session scheduled for 30 minutes (with the possibility to expand time if necessary).
Informed consent was obtained in advance, and interview questions were framed around five high-level categories to get a better sense of rural educators' daily responsibilities and challenges, the disruptions or improvements they have experienced since GenAI's introduction, and their visions for how it could support students and classrooms.
Following Kallio et al.'s ~\cite{kallio2016systematic} framework for crafting an interview protocol, our interview questions were iteratively developed by reviewing related literature (e.g., ~\cite{hardy2019rural, jasanoff2015future, walsh2012putting, tan2024more, harvey2025don}), drawing on the authors' collective expertise (see \S\ref{sec:position}), and piloting within the research team. 
This process ensured our interview questions were relevant in helping us address our research questions (see \S\ref{sec:intro}) as well as foregrounding
educators' lived experiences while also retaining flexibility inherent to a semi-structured approach. 
We provide example guiding questions for each of the five high-level categories below.
\begin{enumerate}
    \item Educator's teaching role and characteristics of their community: To understand how rural environments shape educators' teaching experiences and student interactions, we examined how educators perceive and describe their communities, allowing them to focus on both challenges and assets. These descriptions helped ground our analysis in how participants define what ``rural'' means in their local contexts and interpret their experiences:
    \begin{itemize}
        \item How would you describe the community where you currently teach?
        \item What traits make you categorize it this way?
    \end{itemize}
    
    \item Typical school day activities: To establish a baseline for understanding educators' workloads, these questions explored daily routines and patterns in teaching. We sought whether GenAI emerged naturally in educator practices:
    \begin{itemize}
        \item How do you interact with and engage your students while teaching?
    \end{itemize}
    
    \item The impact of GenAI on teaching responsibilities: To understand how GenAI might reshape teaching practices, questions focused on potential enhancements and disruptions, probing these impacts explicitly when not previously addressed and situating them within educators' rural community descriptions:
    \begin{itemize}
        \item How do you use Generative AI in your teaching (if at all), and what tasks do you choose to use or not use it for?
    \end{itemize}
    
    \item Speculation on GenAI tools that could alleviate rural educators' challenges: To encourage educators to reflect on challenges specific to them and reimagine ways GenAI tools could address gaps or limitations given their current resources:
    \begin{itemize}
        \item How would this Generative AI-based tool change your role as a teacher in this activity? 
        \item How do you think the tool would affect student engagement?
    \end{itemize}
    
    \item Concerns regarding GenAI: To gauge educator's perceptions about risks and limitations of using GenAI in their classrooms and rural communities:
    \begin{itemize}
        \item What concerns do you have about using Generative AI as part of your teaching curriculum? 
        \item What additional support or resources do you need to effectively integrate Generative AI into your teaching? 
    \end{itemize}
\end{enumerate}

The interviewer relied on this protocol as a guideline, and probed and adapted the questions based on individual responses.
This semi-structured approach helped capture unique perspectives that predetermined questions might have missed.

\subsection{Qualitative Data Analysis Methodology}
\label{sec:data-analysis-method}
\subsubsection{Interview Data Analysis}
Interviews were recorded and transcribed via Zoom, with transcripts thoroughly cleaned and de-identified prior to analysis. 
With a grounded theory-informed approach ~\cite{glaser2017discovery, kushner2003grounded}, we began with open coding the interview data, allowing codes and themes to emerge inductively from participants’ responses. 
The first three authors independently coded the transcripts to develop an initial codebook. 
They then met to discuss and reconcile coding discrepancies, merge overlapping codes, and identify overarching themes. 
Using the digital whiteboard tool Miro\footnote{\url{https://miro.com}} for affinity diagramming, they iteratively clustered related codes into broader thematic categories.
Following the initial theme identification, they met again to resolve any disagreements and refine the final set of themes emerged across the interviews. 
All revisions were discussed and agreed upon by all researchers, resulting in 31 low-level codes organized into 11 high-level codes (see~\autoref{fig:codebook}).
Inter-rater reliability was not calculated, as the objective of this analysis was to identify emergent themes rather than achieve coding agreement, following the recommendations of McDonald et al. ~\cite{mcdonald2019reliability}.

\subsubsection{Survey Data Analysis}
\begin{figure*}[t!]
    \centering
    \includegraphics[width=\linewidth]{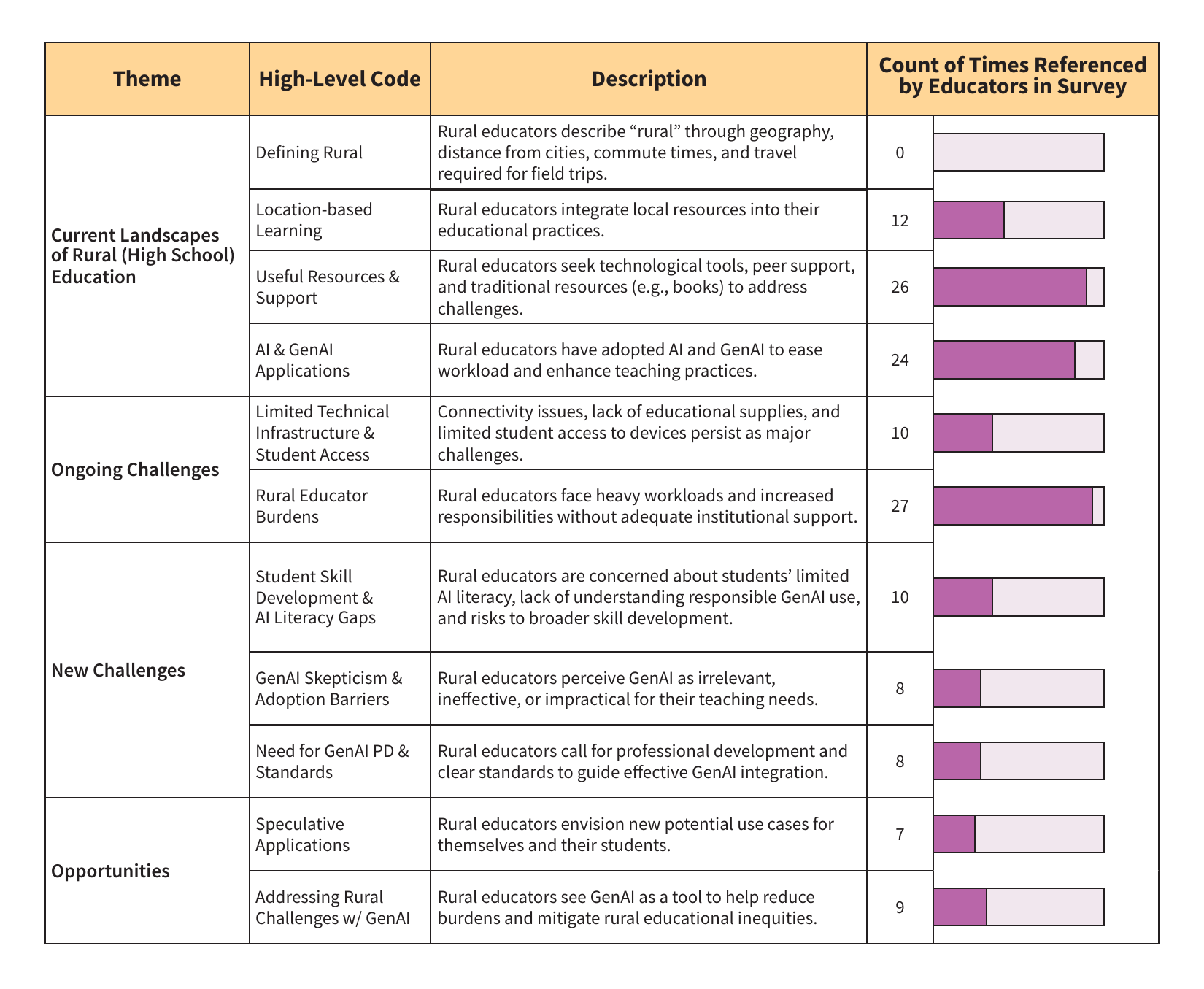}
    \caption{Our codebook was developed by deductively coding the interview data. This codebook was then used to code the open-ended survey questions with the same nuanced context. We counted the number of times these codes were referred to in the survey open-ended responses, which is represented by bar charts.}
    \label{fig:codebook}
    \Description{A table of our codebook developed from our interview data analysis and applied to analyze the open-ended survey questions. It details our four themes, 11 high-level codes, description of each code, and counts of each code referenced by participants in the survey, visualized by bar charts.}
\end{figure*}

To analyze the open-ended survey questions, we conducted a deductive qualitative coding analysis~\cite{fife2024deductive} using the codebook (\autoref{fig:codebook}) developed for the interview analysis, then quantified the prevalence of each code. 
We analyzed the interview data first because the interviews captured richer narratives of rural educators' lived experiences, which served as a strong foundation for identifying emerging themes. 
We then applied this established codebook to the survey responses, ensuring shorter, less-detailed answers were interpreted within the same nuanced context as the interview data.
This allowed us to identify alignment and convergence across both parts of the study.
A similar method has been used in prior research ~\cite{dempsey2018handling}, where qualitative interviews were first analyzed to develop a coding framework, which was then applied to survey data to inform a broader interpretation.

The first author independently coded the open-ended responses, discussed the categories with the other authors, and refined the coding.
They then counted each code's occurrences as seen in ~\autoref{fig:codebook} (right-most column).
As a reminder, survey responses from participants who failed either of the attention checks were eliminated from further analysis.

\subsubsection{Theory-informed Reanalysis}
To deepen our interpretation, the first author then engaged in a theory-informed reanalysis, drawing on critical rural theory~\cite{thomas2013critical} as an analytic lens. 
This secondary pass enabled us to connect emergent themes to broader structures of power, equity, and agency shaping rural educators’ experiences with GenAI.
Our approach aligns with Alfred Schutz’s social phenomenology~\cite{schutz1967phenomenology}, which emphasizes understanding participants’ lived experiences before situating them within broader social and theoretical contexts.
Similarly, Proudfoot~\cite{proudfoot2023inductive} proposed an inductive–deductive hybrid approach for mixed-methods research, in which interview and survey data are first analyzed inductively, and then interpreted deductively through a guiding theoretical framework\footnote{Note that this secondary-analysis did not influence or change the themes and the developed codebook; rather, it provided a lens for us to interpret and understand the findings from the primary-analysis.}.
The results from this secondary analysis is presented in \S\ref{sec:discussion}.

\section{Author Positionality}
\label{sec:position}
We reflect on our positionalities and present our backgrounds to contextualize our experiences and demonstrate our credibility in studying these populations, while recognizing potential biases ~\cite{holmes2020researcher}. 
The first and fourth authors both grew up in rural areas.
The other authors, while not from rural communities, bring relevant experience through their prior research and engagement with rural education.
The second, third, and fourth authors have conducted prior research on rural education and communities, and they work with K-12 students from these communities to empower them with access to technology and literacy.
Additionally, the first, fifth, and last authors focus on responsible AI and ethics research serving underrepresented communities with implications for more equitable futures.
We acknowledge our research team is privileged in having access to resources and networks, given our professional and academic backgrounds, which may not reflect the lived realities of our participants.
We finally emphasize as a research team, we approach AI and GenAI from a pragmatic stance, acknowledging both its transformative possibilities and potential harms.
Rather than adopting either utopian or dystopian perspectives, we focus on studying ways to empower users to engage responsibly, such as through thoughtful intervention and design.
This balanced perspective inevitably shaped our study design, (e.g., the questions posed to the participants), analysis, and guides the narrative of this paper. With these considerations, we have placed the voices of rural educators at the center of our work.

\section{Results} 
\label{sec:results}

In this section, we present our survey and interview results collectively, focusing on rural educators' lived experiences. 
Since we used the same codebook for both data sources (see \S\ref{sec:data-analysis-method}), we present the findings together rather than separately.

We begin by contextualizing the rural educational landscape as defined by the participants (\S\ref{sec:current-landscapes}).
These findings helped establish an understanding of rurality and describe their everyday teaching, student learning, and measures of success.
While these patterns align with prior research~\cite{corcoran2009instruction, rosenshine2012principles, cooper2014eliciting, blazar2017teacher, allen2013observations} and primarily serve to ground the rest of participant responses, we present the full results in the supplemental material.


Building on this foundation, we next examine persisting challenges (\S\ref{sec:ongoing-challenges}), identify emerging new challenges (\S\ref{sec:new-challenges}), and highlight opportunities that GenAI\footnote{In this paper, we use the term ``GenAI'' to refer specifically to generative artificial intelligence applications. However, during interviews and surveys, participants frequently used the term ``AI'' more generally. Where relevant, we retain participants’ terminology while situating it within the broader context of GenAI's impact on rural education.} presents for rural high schools (\S\ref{sec:opportunities}).
These results describe participants' lived experiences as they were expressed.
We then interpret our findings through a critical rural theory lens~\cite{thomas2013critical} for a deeper theory-informed analysis, in the discussion section (\S\ref{sec:discussion}).

In addition to the qualitative analyses of the interviews and the open-ended responses from the survey, we analyzed the multiple-choice questions from the survey by reporting counts and percentages, where relevant, to complement our qualitative findings.
Initially, we analyzed participants' responses to the multiple-choice questions by separating them by state.
We hypothesized this separation could be meaningful since rural regions within each state might face unique challenges.
However, educators were largely aligned in their responses across the states. 
While this analysis and distributions of responses are available in the supplementary material, here we present our results \textit{without} state-based separation.

For the remainder of this paper, we refer to participants with unique identifiers, with  \textbf{S\#} for survey participants and \textbf{I\#} for interviewees.
Educators who participated in both studies are referenced uniquely by both their identifiers depending on whether we are discussing their survey or interview responses.
As a reminder, the study included 31 unique participants: 4 who completed both the survey and interview, 2 interview-only, and 25 survey-only participants.

\subsection{Current Rural Educational Landscape}
\label{sec:current-landscapes}
To ground our findings, we gain an understanding of the current landscape in rural communities where participants defined ``rural'' in their own words and shared current educational practices and tools they employ.
\\\\
\noindent\textbf{Defining Rural.} 
When describing what makes their communities rural, many participants emphasized the geographic location.
\textbf{I6} characterized the area as \textit{``mainly farmland and farms''}, while \textbf{I5} described it as a region with \textit{``large chicken, tobacco, soy, and cotton farming''}.
Similarly, \textbf{I2} shared their community as an \textit{``itty-bitty town''}, where local activities are largely limited to hiking in the nearby mountains.
Several educators noted their communities were in the \textit{``middle of nowhere''} (\textbf{I1}) with the nearest city often an hour or two away (\textbf{I3}).
In some counties, there are no cities and driving to the county line can take 30 minutes (\textbf{I6}).
The schools were described as ``standalone'', where many are the only high school in their counties (\textbf{I5, I6}).
\textbf{I1} stressed how small each class is, sharing the most recent graduating class had only 35 students.

Traveling to and from school takes significant time.
Student commutes range from 10 minutes to over an hour by bus (\textbf{I3, I4}), and in some districts, such as \textbf{I5}'s, most educators drive in from surrounding counties.
Field trips are additionally limited.
For instance, \textbf{I4} described how their class takes trips to regional coastal resources to see oyster beds in the area.
Some destinations require at least an hour of travel (\textbf{I2}) and other field trips are as far as four or five hours away (\textbf{I1}).
Given the cost of transportation, some schools avoid field trips entirely, with \textbf{I5} noting \textit{``it's a lot of gas to get anywhere from here''.}
To provide students with broader experiences outside their community, some educators have opted for virtual field trips as an alternative (\textbf{I1}).
\\\\
\noindent\textbf{Location-based Learning.} 
Rural high school educators are leveraging and integrating local resources into their educational practices to bridge learning with everyday life and students' potential future careers.
Outdoor STEM activities range from tree planting, field investigations, and environmental protection projects (\textbf{S3}) to experiential learning through outdoor labs that connect classroom concepts to real-world applications (\textbf{S5, S7, S17}).
Agriculture and farming are also central to learning for many rural communities. 
Some participants teach agriculture more broadly (\textbf{S9, S10}) while others specialize in teaching plant growth and sustainability (\textbf{S25, S34}) or animal science in breeds, biotechnology, and animal care (\textbf{S24}).
Additionally, hands-on skills are prioritized where students learn about plant propagation, electrical wiring, and welding (\textbf{S26}).
Local industries and culture influence rural education.
For example, \textbf{I4} described visits from the local marine center to their school and \textbf{I5} shared nearby farms brought live chickens for students to see.
Similarly, \textbf{I1} shared how local legends inspire creative projects: \textit{``...living in Arizona, we have legends, and the legend of the lost Dutchman's mine is one that always peaks people's curiosity. And so I created a lab group project...[using] all these legends and clues to find where you think the lost Dutchman's mine is...''} 
\\\\
\noindent\textbf{Useful Resources and Support.}
The majority of survey participants indicated a preference for seeking technological and technical resources to address challenges in their classrooms.
24.1\% of participants said they rely on general technology like smart boards, projectors, and other presentation tools to support instruction.
31\% of participants mentioned online platforms, learning management systems, and educational apps are commonly used to provide supplementary materials, interactive exercises, and organizational support for both educators and their students.
Participants turn to internet resources and online content providers including blogs, \textit{YouTube\footnote{\url{https://youtube.com}}}, \textit{code.org\footnote{\url{https://code.org}}}, \textit{Edpuzzle\footnote{\url{https://edpuzzle.com}}}, and \textit{Teachers Pay Teachers\footnote{\url{https://teacherspayteachers.com}}} (\textbf{S17, S21, S23}) for new lesson ideas and plans.
While technology-based resources are more sought out, educators still find value in human-based support, coming from fellow educators (\textbf{S2, S24}), students (\textbf{S4}), parents (\textbf{S29}), administrators (\textbf{S24}), and professional online communities dedicated to teaching, such as forums and social media groups (\textbf{S10, S26}).
Additionally, traditional resources continue to play a critical role, as educators learn about valuable theoretical foundations and best practices from textbooks (\textbf{S10, S12}), educational research papers (\textbf{S10}), books (\textbf{S29}), and technical journals (\textbf{S26}).
\\\\
\noindent\textbf{AI and GenAI Applications in Rural High Schools.} 
In addition to general technology-based resources, AI and GenAI are increasingly being adopted by educators as tools to ease their workload and enhance their teaching practices.
A large majority ($80.6\%$) of participants shared they use AI and GenAI primarily as support tools for themselves.
There are also shifts in teaching responsibilities, such as educators using GenAI to help with their daily activities and lesson preparation (\textbf{I3, I4}).
Among participants who use GenAI, nearly half ($48.7\%$) identified their most common applications as curriculum development, grading, differentiating instruction, and generating personalized materials such as student worksheets, lesson plans, rubrics, and syllabi.
These participants emphasized GenAI's ability to save time was especially valuable in generating quick responses tailored to their needs.
Others noted GenAI helped provide ideas and agendas for after-school activities (\textbf{S19, S22}), assisted with general writing (\textbf{S8, S17, S21, S22, S26)}, and summarized texts and data (\textbf{S19, S22}).
Differentiating instruction stood out as a particularly meaningful benefit for participants like \textbf{I4} who shared:
\begin{quote}
    \textit{``...I had a kid who was having a really tough day, really stressed out by writing... And [the student says] `... I'm having trouble today. [I] really don't think I can do this'.... I remember asking like, `Okay, if I made these two or three changes to it. Would that help?' And they said, `Yes, it would'. And normally I'm able to make those changes myself. But I would need time outside of class... probably take me 25 to 30 minutes to put together... Within, you know, a couple of minutes I had fed that back into [ChatGPT]... and it spit it right back out. And I had it. That kid had what they needed instantly.''}
\end{quote}

Participants reported using a variety of GenAI-based systems.
Platforms included \textit{ChatGPT\footnote{\url{https://chatgpt.com}}} ($40\%$), \textit{Gemini\footnote{\url{https://gemini.google.com}}} ($13.3\%$), \textit{Perplexity\footnote{\url{https://perplexity.ai}}} ($20\%$), and \textit{Claude\footnote{\url{https://claude.ai}}} ($26.7\%$).
Aside from these generic tools, some participants reported utilizing educator-specific GenAI systems, such as \textit{Gamma\footnote{\url{https://gamma.app}}} ($6.7\%$), \textit{Magic School AI\footnote{\url{https://magicschool.aid}}} ($33.3\%$), \textit{Brisk\footnote{\url{https://briskteaching.com}}} ($26.7\%$), \textit{SchoolAI\footnote{\url{https://schoolai.com}}} ($13.3\%$), and \textit{Diffit\footnote{\url{https://web.diffit.me}}} ($6.7\%$).

\subsection{Ongoing Challenges}
\label{sec:ongoing-challenges}

\begin{figure*}[t!]
    \centering
    \begin{subfigure}[t]{0.89\textwidth}
        \centering        
        \includegraphics[width=\textwidth]{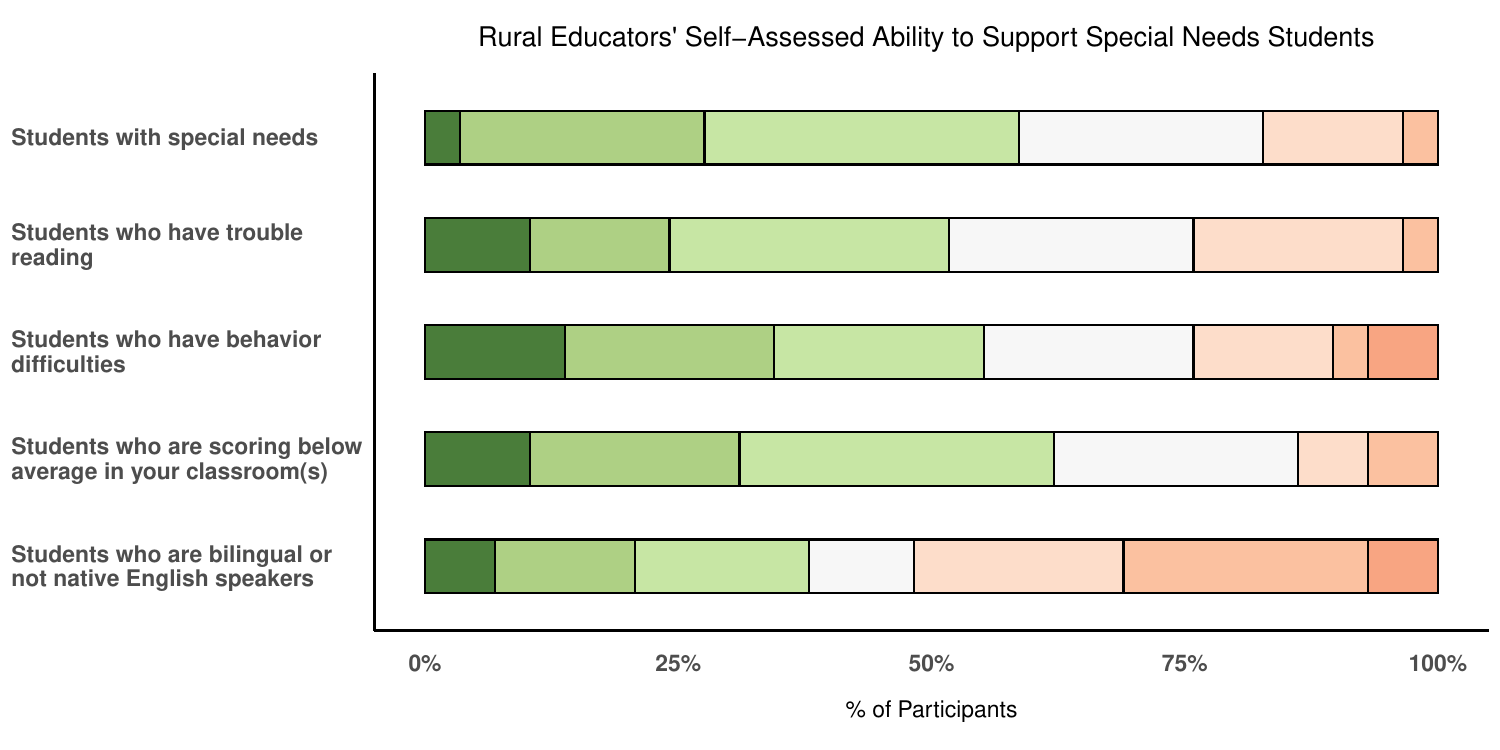}
    \end{subfigure}%
    \begin{subfigure}[t]{0.1\textwidth}
        \centering
        \includegraphics[width=\textwidth]{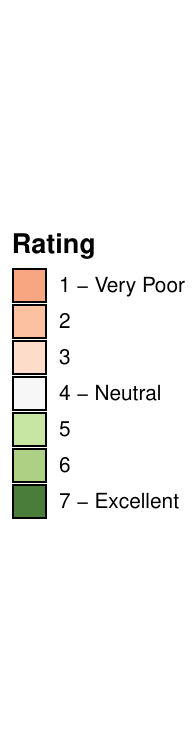}
    \end{subfigure}
    \vspace{-.5cm}
    \caption{Participants' distribution of responses, measured on a 7-point Likert scale, on their ability to support students with diverse needs based on current resources and support from their schools.}
    \Description{The stacked bar chart details the distributions of participants' responses assessing their ability to support students with diverse needs based on current resources from their schools. Ratings were on a 7-point Likert scale from ``Very poor'' to ``Excellent''. The majority of participants reported satisfactory abilities across all categories.}
    \label{fig:ability-rate}
\end{figure*}

\noindent\textbf{Limited Technical Infrastructure and Student Access.}
Although GenAI has started to show promise, rural schools continue to face persistent barriers that limit its impact.
A major concern voiced by 6 participants from the survey is technical infrastructure, which includes unreliable internet, insufficient electronic and hardware devices, and subscription costs.
Without these, classrooms are prevented from fully adopting GenAI as an educational tool or using it more for classroom support.
Besides connectivity, participants like \textbf{S12}, \textbf{S13}, and \textbf{I6}  described how difficult it is to obtain educational supplies and materials for their classrooms.
\textbf{I6} described a sense of \textit{educational abandonment} due to their school's location, reflecting, ``\textit{I don't really see a lot of professional development (PD) opportunities unless it's like...[over Zoom] and stuff like that. We just don't have that. Nobody's going to come to [my county]}.'' 
On the other hand, several participants noted the inequities in student access, where
many of them lack internet access or personal devices at home (\textbf{S6, S24, S25}).
\textbf{S24} and \textbf{S25} reflected similar sentiments that while GenAI could \textit{``allow [students] to look at the world in a whole new light''}(\textbf{S24}), the benefits of GenAI remain out of reach for many classrooms until these infrastructure and resource gaps are addressed.
In some cases, schools attempted temporary solutions.
For instance, \textbf{I3} described a now-discontinued program where their school used to offer \textit{``need-based laptop library loans''} and subsidized purchase options for students.
\\\\
\noindent\textbf{Rural Educator Burdens.}
There are already a limited number of \textit{``expert educators''} (\textbf{S12}) in rural schools, particularly in STEM fields ~\cite{rude2018policy}, and as a result, educators have faced increased responsibilities to teach multiple grades and subjects based on need.
\textbf{I1} acknowledged this challenge, noting a broader \textit{``teacher problem''} where educators \textit{``are being asked to do things that they normally aren't able to''}.
For example, some participants reported they are the only teacher at their school that teaches science (\textbf{I1}) and math (\textbf{I6}), and others described teaching across middle and high school levels (\textbf{I2}) as well as teaching both regular and honors classes in a single classroom (\textbf{I5}).
Despite these constraints, most participants rated their ability to support students across a range of diverse needs as sufficient, as seen in \autoref{fig:ability-rate}.
However, they noted areas where they felt less equipped, particularly in supporting bilingual students or those who are not native English speakers ($37.9\%$).
While educators are managing with their current resources and support, they are challenged with high teaching staff turnover.
Educators like \textbf{S5} cautioned about having a \textit{``solid core of veteran teachers..., however, we will lose those teachers within the next 3 to 5 years due to retirement. I think it may be challenging for new teachers to meet the needs...''}.

A need shared by all participants is greater support for their teaching responsibilities.
Participants from the interviews identified needs for assistance with curriculum differentiation (\textbf{I2}), resources for creating fun activities (\textbf{I1}), time-saving strategies for grading (\textbf{I2}), and stronger mechanisms for student support (\textbf{I6}).
In addition, many repetitive tasks consume participants' time, which take away their ability to focus on instruction and student engagement.
For survey participants, these included grading ($34.5\%$), lesson planning and preparation ($51.7\%$), administrative work like attendance and emails ($44.8\%$), and classroom management with student behavior ($20.7\%$).

\subsection{New Challenges}
\label{sec:new-challenges}


\noindent\textbf{Student Skill Development Strains \& AI Literacy Gaps.}
A major concern voiced by participants was students’ limited AI literacy and lack of understanding of responsible GenAI use.
Without explicit guidance or oversight, these educators feared students were experimenting with these tools in ways that exceeded classroom parameters and expectations (\textbf{I1, I3, I4}).
One participant stated the inevitable with many new mainstream technologies: ``\textit{You know the kids, they're going to experiment with this stuff. They're going to go outside the parameters you set for them. They're going to do things with tools that you don't expect...}'' (\textbf{I4}).
Such experimentation could fuel students' belief in the infallibility of GenAI, leading to an opportunity to allow, as per \textbf{S9}, \textit{``academic dishonesty''}.
$31\%$ of participants worried GenAI will inhibit student autonomy from developing their own academic skills, lose their creativity, and their students will see GenAI as a shortcut to bypass learning, especially for writing.
\textbf{S34} noted students who lack motivation, struggle with writing, or fail to see value may be more likely to submit AI-generated writing as their own.
In addition, participants believed overreliance on GenAI for tasks like writing can worsen knowledge gaps, especially for those students who have a limited understanding of plagiarism (\textbf{S19}) and do not know how to properly cite sources (\textbf{S6}).
These challenges are further exacerbated by the lingering effects of the COVID-19 pandemic as \textit{``students are still behind''} academically (\textbf{S7}, \textbf{S6}).

Participants also shared a common fear GenAI usage could diminish essential life skills, such as building human relationships and interpersonal skills.
Specifically, some participants like \textbf{S10} worry about extensive use of GenAI during classes, as
\textit{
``...it might reduce the level of direct interaction and engagement between the teacher and students''}. 
As a result, GenAI could potentially affect students' learning of life skills through hands-on activities, discussions, and guided exploration that foster deeper understanding and interest in science (\textbf{S10}).
Similarly, \textbf{I1} worried, \textit{``I don't want this next generation to come up to have a huge disconnect because they have a personal relationship with Siri and not the person sitting next to [them].''}
Beyond relationships, educators raised the importance of teaching beyond textbooks (\textbf{I1}, \textbf{S8}) and how GenAI can have an indirect negative impact on students' futures.
As \textbf{I1} explained, \textit{``...We try to prepare them for life. And a lot of students I have...their home life isn't great and so we're filling in those gaps, trying to give them as much as we can...''}.
\textbf{I1} further cautioned, \textit{``I'm like, I want you to be the best that you can be, and if you are given a crutch [in your] learning formative years, you're never going to be able to walk on your own''}.
\\\\
\noindent\textbf{GenAI Skepticism \& Adoption Barriers.}
Although many participants shared how they use GenAI in their classrooms (see  \S\ref{sec:current-landscapes}), others viewed these technologies as irrelevant or ineffective for their teaching needs.
For instance, a few participants reported they \textit{``don't have a need for it at this time''} (\textbf{S5}), did not find it valuable (\textbf{S16}), or were uncertain on how GenAI could be applied in their subject areas (\textbf{I6}).
Others felt GenAI falls short in its ability to \textit{``perfectly convey''} the content educators want students to learn (\textbf{S29}).
In particular, \textbf{S29} believed the current AI search base is \textit{``too cluttered to provide safe and reliable teaching information''}.
Educators like \textbf{S24} questioned GenAI's effectiveness and relevance within rural schools, relating back to their school's lack of access to technology.
However, \textbf{I5} cited this ``\textit{rural delay}'' could be seen as a temporary advantage rather than a disadvantage: \begin{quote}
\textit{``...I read different articles online from different [non-rural] school systems in different states and the struggles they're having and the mandates that are being put in place. I see...the struggles they're having, and I'm like oof, we're such a small school that right now that hasn't hit us. We're not there yet. It hasn't gotten to us yet. So it's not an issue yet...''}
\end{quote}

Some participants maintained that GenAI should remain a tool for educators, not students (\textbf{S1, S14}).
A few others, like \textbf{S10}, expressed concern about their own reliance on GenAI which in turn could \textit{``hinder students' development of their own problem-solving and critical thinking abilities''}.
In addition to individual concerns, educators like \textbf{I3} are concerned about technological resistance and lack of acceptance from older, non-technical educators in their school who have developed a ``\textit{necessary struggle attitude}'' for students to succeed:
\begin{quote}
\textit{``I worry that the reaction of the people who feel very uncomfortable with many of the [GenAI] concepts that we've been describing, it's gonna color and ... affect the access that my young students have to this technology in a way that ... severely disadvantages them. [When] they enter the ... world of adulthood, and an era in which [these] tools are even more ubiquitous ..., if they don't know how to use the tools, or they don't understand how tools work in a general sense...''}.
\end{quote}
\textbf{I5} resonates with these feelings with their ``\textit{old school mind}'' expressing, ``\textit{I'm sure there's so much out there that I don't [know] and it gets overwhelming. And I'm like, `Oh, shut it down!'...}''
\textbf{I5} has also confessed they reverted their students back to ``\textit{a lot of pencil and paper writing}'' due to their own expressed aversion towards GenAI since its emergence.
However, from the CARS questionnaire (see \autoref{fig:tam-cars}), these anxieties are not limited to older or non-technical educators. 
A majority of participants from the survey reported low confidence in their ability to learn GenAI skills that could enhance their teaching ($79.3\%$) and difficulties keeping up with new technologies and applying them in the classroom ($75.9\%$).
\\\\
\noindent\textbf{Need for GenAI Professional Development \& Standards.}
With the limitations GenAI brings about, participants unanimously expressed concerns about their own lack of AI literacy needed to teach responsible and safe usage of these tools. 
After finding out one of their students used ChatGPT for an assignment, \textbf{I2} wanted to teach their class about responsible use, however, ``\textit{...I didn't have all the resources needed to be able to teach them a full lesson on the appropriate level of when and when to not use it, or how to use it}''.
Not only do educators not have the resources, \textbf{S20} added \textit{``many teachers don't feel comfortable leading those discussions because they don't know how to use [or] are not comfortable with AI [themselves]''}.
Although GenAI is recognized as a \textit{``game-changer''} (\textbf{S25}) in education, many participants emphasized the need for receiving training and Professional Development (PD) on GenAI to empower them in teaching more effectively and guiding their students when and how to use these tools (\textbf{S14, S18, S20, S25}); especially, with more emphasis on \textit{recurring and up-to-date} PD and trainings to keep pace with technological changes (\textbf{S18}).
With proper training, educators could then \textit{teach} and \textit{guide} students not to use GenAI tools to facilitate cheating, discuss the (implicit) plagiarism associated with overreliance, and help position GenAI as rather a \textit{``thought partner''} (\textbf{S21}) instead.
This will also empower them to teach students critical thinking skills for GenAI use, including awareness of shortcomings and harms such as content hallucination and misinformation (\textbf{S18, S23}).

Participants pointed to the absence of clear standards and guidelines for GenAI use within K-12 and other educational settings.
Currently, educators have the new added responsibility of detecting AI-generated content, but many do so without confidence in distinguishing it from genuine student work (\textbf{I1, I2, I5}).
Some educators, like \textbf{I6}, shared they are waiting for direction from their state or district before integrating GenAI into their teaching.
At the same time, educators acknowledge even with clear standards and guidelines, \textit{``... there's always that one parent...that would have a problem with me using an AI...''} (\textbf{I1}).
This comment highlights broader fear and uncertainty about AI use among other stakeholders in a student's life, which underscores the need to standardize responsible AI policies and practices in K-12 education to ensure community-wide buy-in and stakeholder awareness of usage guidelines.
We will describe this and other opportunities and needs in the following section.

\subsection{Opportunities}
\label{sec:opportunities}

\begin{figure*}[t!]
    \centering
    \begin{subfigure}{0.44\textwidth}
        \centering        
        \includegraphics[width=.9\textwidth]{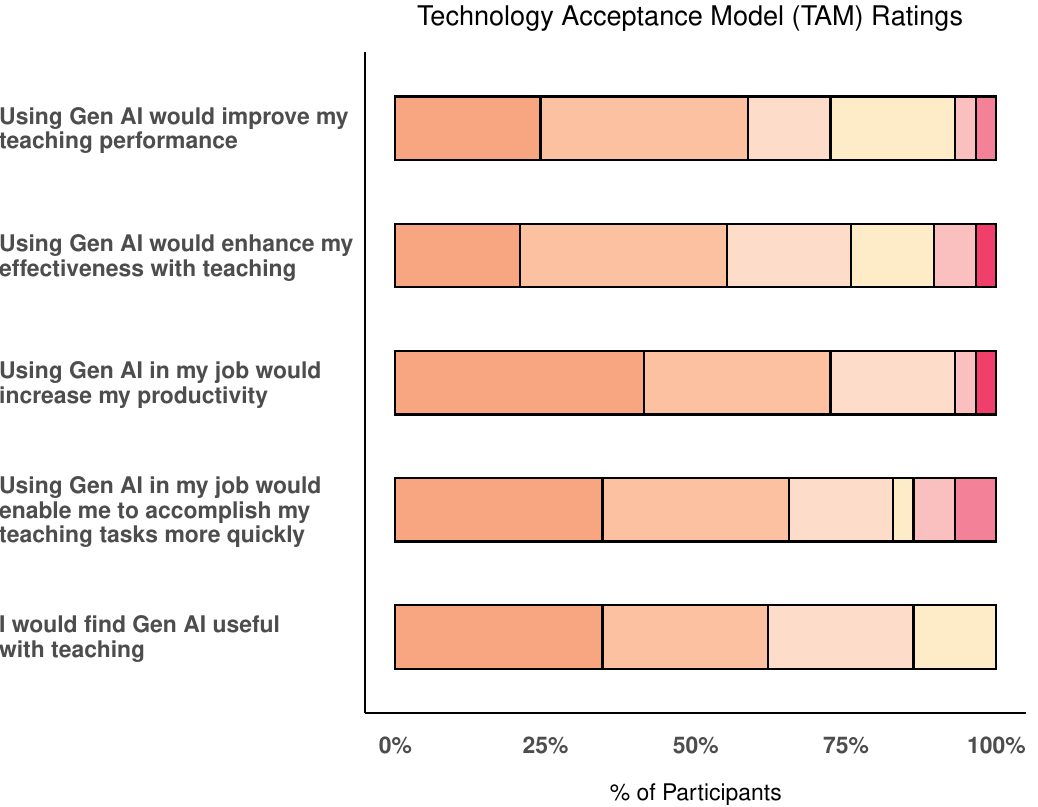}
    \end{subfigure}%
    \begin{subfigure}{0.44\textwidth}
        \centering        
        \includegraphics[width=.9\textwidth]{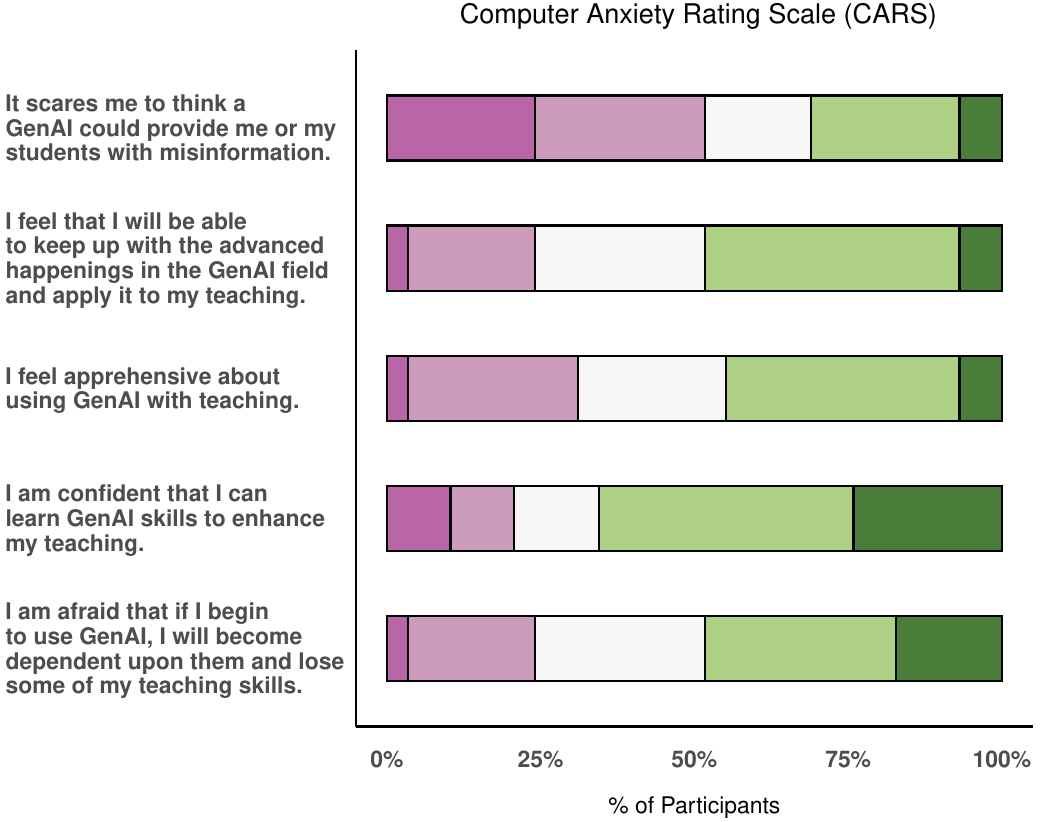}
    \end{subfigure}%
    \begin{subfigure}{0.108\textwidth}
        \centering
        \includegraphics[width=\textwidth]{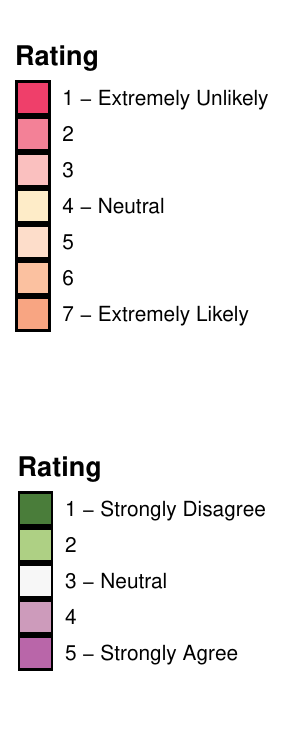}
    \end{subfigure}
    \caption{Distribution of responses to our modified Technology Acceptance Model (TAM) (left) ~\cite{davis1989perceived} and Computer Anxiety Rating Scale (CARS) (right) ~\cite{heinssen1987assessing} questionnaires. Participants rated their perceptions and attitudes on  7-point and 5-point Likert scales, respectively.}
    \Description{The left visualization is a stacked bar chart that captures the distributions of responses to our modified Technology Acceptance Model (TAM) questionnaire from the online survey. Responses were on a 7-point Likert scale from ``Extremely Unlikely'' to ``Extremely Likely''. The majority of participants perceived generative AI favorably.
    On the right, the stacked bar chart visualizes the distributions of responses from our modified Computer Anxiety Rating Scale (CARS) questionnaire from the online survey. Attitudes were measured on a 5-point Likert scale from ``Strongly Disagree'' to ``Strongly Agree'' and the majority of participants shared they had anxieties towards generative AI.}
    \label{fig:tam-cars}
\end{figure*}

Despite the challenges GenAI introduces and exacerbates within the context of rural educational settings, our participants remain hopeful about its potential. 
Some stated they have already started to experiment with incorporating GenAI in their classroom,
such as \textbf{I4}, who is using GenAI to provide feedback and support to students for writing assignments.
Even those who described themselves as novices (\textbf{S13}) or who do not currently use GenAI have expressed optimism about GenAI's potential.
For instance, \textbf{I5} and \textbf{I6} try to keep an open mind and actively seek opportunities to understand how to integrate this technology into their classroom and lesson-planning practices.
Directly relevant to these positive perceptions, from the TAM questionnaire, the majority of participants from the survey highly rated GenAI's ability to help them accomplish teaching tasks more quickly ($82.8\%$), improve teaching performance ($72.4\%$), increase productivity ($93.1\%$), enhance effectiveness in teaching ($75.9\%$), and demonstrate broad usefulness ($86.2\%$) (see \autoref{fig:tam-cars}).
These positive views are consistent with participants' actual GenAI usage patterns, documented in \S\ref{sec:current-landscapes}.

Beyond these favorable attitudes, participants acknowledged the importance of GenAI, emphasizing educational settings must adapt and embrace it accordingly as it is ``\textit{here to stay}'' (\textbf{I5}).
For some, GenAI-supported tools are here to provide support, but not replace authentic work for both students and teachers (\textbf{I3}).
Several educators contextualized GenAI emergence by drawing parallels to historical technology transitions and their impacts on education, namely the introduction of the internet (\textbf{I1, I5, I6}).
Similar to these tools, some participants believed increased GenAI usage leads to improved ability to leverage it better (\textbf{S7, S13}).
Echoing this sentiment, \textbf{I4} emphasized there will be both learning and acceptance curves, and concerns will be alleviated with time, elaborating, 
\begin{quote}
\textit{`` ... It's going to be a part of our educational system, likely a part of the way we teach forever moving forward. And even more likely that students [will be] adopting it. And we will see a higher and higher percentage of students using it as the years go on.''}
\end{quote}

Given that there are significant uncertainties that exist around how to facilitate this transition in rural educational settings, we have identified three high-level themes regarding promising GenAI opportunities in rural schools, drawn from speculative questions with survey participants and extended discussions with interviewees:
\begin{figure*}
    \centering
    \includegraphics[width=1\linewidth]{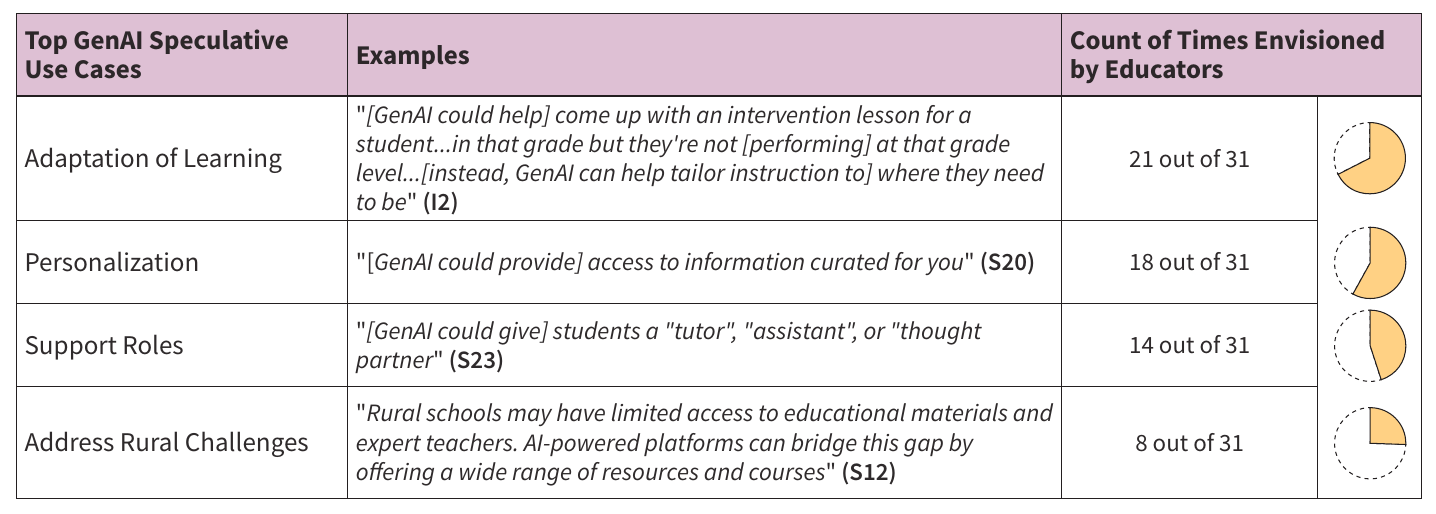}
    \caption{The top four speculative GenAI use cases envisioned by participants for themselves and their students. These use cases drew on applications identified in non-rural contexts ~\cite{abolnejadian2024leveraging, belghith2024testing, han2024teachers, oh2024exploring, shahriar2023and, tan2024more, taneja2024jill, yang2023pair}, survey open-ended responses, and interview insights. We include the number of participants who agreed with each use case.}
    \label{fig:speculations}
    \Description{A table of providing an overview of the top four generative AI speculative use cases participants envisioned for themselves and their students. Each use case has corresponding example quotes drawn from interviews and survey open-ended responses. In addition, the number participants who agreed with each vision is enumerated as a proportion and visualized as a pie graph.}
\end{figure*}
\\\\ 
\noindent\textbf{Speculative Applications.}
In the survey, when presented with potential GenAI use cases for educators and their students---based on applications that are already employed or studied in non-rural areas---the majority of participants saw potential for themselves ($96.6\%$) and their students ($86.2\%$) (see~\autoref{fig:speculations}).
For educators, the most widely supported GenAI uses were automating admin tasks ($53.6\%$), creating adaptable teaching material ($82.1\%$) and personalized content ($64.3\%$), generating questions for quizzes ($71.4\%$), and functioning as a teaching assistant ($60.7\%$).
For students, most popular applications include GenAI as a homework support system ($72\%$), ideation tool ($64\%$), and generator for personalized exercise activities ($68\%$).
Participants were enthusiastic to improve existing rural educational processes and student engagement (\textbf{S3, S25}).
Our discussions with educators during the interviews expanded on these visions and speculative designs, with ideas spanning both new educational applications and enhancements to existing GenAI tools that could support these applications.

Participants envisioned new GenAI tools could support them in curriculum planning and design, specifically by allowing them to adapt curricula to different grade levels and students' skill levels on demand.
For example, \textbf{I2}, who teaches both middle and high school STEM and humanities classes, expressed a need for GenAI outputs that are less ``professionally worded'' and require fewer follow-up adjustments, so that lesson materials better align with their students’ reading levels and grade-specific needs.
Some participants wanted stronger academic integrity tools, such as more reliable cheating detection and mechanisms for educator oversight through custom parameters (\textbf{I1, I4, I5, I6}).
For students, unique GenAI applications could act as \textit{`peer-like'} (\textbf{I4}) writing partners with awareness of the intended audience (\textbf{I3}), and assist with generating questions from multimedia sources (\textbf{I1}). 
While these uses may not be necessarily new applications, these participants were not previously aware current GenAI systems were capable of such applications.
Some educators noted achieving their desired outcomes often required using multiple GenAI tools, and expressed a preference for a single system capable of handling all tasks more efficiently.
\\\\
\noindent\textbf{Addressing Rural Challenges with GenAI}
Given the existing challenges rural communities continue to face, where access to resources and educators is often limited, participants see opportunities for GenAI to reduce these burdens (\textbf{I2}) and reduce educational resource inequalities (\textbf{S12}).
For example, GenAI can help bridge the gap by \textit{``offering a wide range of resources and courses''} (\textbf{S12}), such as providing an abundance of high-quality content, advanced educational materials, and learning tools to enhance classroom interactions (\textbf{S3, S12, S33}), which can level the playing field for rural students.
GenAI supports location-based learning by presenting students with opportunities to brainstorm and help \textit{``create projects from resources around them''} (\textbf{S13}).
For students with diverse needs, GenAI can provide \textit{``supports as-needed, at almost any time''} (\textbf{S18, S19}) and act as a \textit{``tutor, assistant, or thought-partner''} that might be unavailable given the financial and regional constraints (\textbf{S23}).
With educators often taking on expanded responsibilities, \textbf{I1} hopes educators can leverage GenAI \textit{``to up [their] game''} to fill instructional gaps.
This vision is shared by \textbf{I6}, who sees GenAI as a valuable subject knowledge recourse.

\section{Discussion}
\label{sec:discussion}

We examine our findings through a \textit{critical rural theory} lens~\cite{thomas2013critical} to discuss how structural, social, economic, and political inequalities favor urban-centered narratives and shape rural communities' experiences with GenAI. 
Given that rural contexts are often overlooked in research, adopting this critical lens is key to understanding how rurality shapes educators' responses in relation to the power and systemic inequities unique to these communities, which other theoretical frameworks (or purely qualitative analysis) might miss. 
We extend this rural-focused discussion by incorporating analytical frameworks from critical feminist theory~\cite{moore2018cat, crenshaw2013demarginalizing} and critical technology studies~\cite{ortolja2022encoding, feenberg2008critical}, examining how \textit{power}, \textit{equity}, and \textit{agency} dynamics specifically influence the interplay between rural educators and GenAI.
By triangulating critical rural theory, prior research, and our own empirical findings, we demonstrate how seemingly neutral GenAI implementation processes reflect embedded power dynamics that can privilege some communities over others and demonstrate how equitable GenAI integration benefits from alignment with unique rural pedagogical contexts, values, and priorities.
We conclude with recommendations for the HCI community to address these disparities (\S\ref{sec:recoms}) and describe opportunities for future research (\S\ref{sec:future}).

\subsection{The Dynamics of Power, Equity, and Agency in GenAI Engagement}
\label{sec:critical}
Rural educational contexts face compounding challenges that intensify GenAI implementation barriers.
As evident from our studies with rural educators and supported by prior work, these communities experience persistent difficulties in recruiting and retaining educators (\textbf{S5})  ~\cite{rude2018policy, jensen2020closing} and many educators approach adopting new technologies cautiously (\textbf{I5, I6}) ~\cite{weeks2020teachers, rude2018policy}.
Additionally, rural students are more likely to come from lower socioeconomic backgrounds ~\cite{lavalley2018out, schaefer2016child, showalter2019rural}, experiencing limited access to reliable internet and technology at home (\textbf{S17}) ~\cite{roberts2023effects}---disparities that became more pronounced during the COVID-19 pandemic (\textbf{S7, I6}) ~\cite{goldberg2021education, roberts2023effects}.
Since technology access has shown to support student learning outcomes ~\cite{fikriyah2022use, demir2022examination}, intentional GenAI integration has the potential to expand opportunities for rural students or, if not implemented thoughtfully, it could further exacerbate disparities.
Here, we analyze how rural educators navigate power and agency in integrating GenAI, and how existing inequities shape these relationships.

\subsubsection{Compounding Power Imbalances: How Ill-defined Definitions Reinforce Urban-Centric Design and Policies}

The power over GenAI integration and \textit{whose} values, knowledge, and lived realities shape its adoption has been largely controlled by urban and suburban technology narratives that assume stable infrastructure and available resources,
a framing that overlooks the complexities of rural educational communities and adaptive strategies they employ under structural constraints.
The exclusions are compounded by a lack of consensus on definitions of ``rural''.
Federal and state agencies apply \textit{urbannormative} classifications that misrepresent rural populations, skew data collection, and mask educational needs ~\cite{goode2012beyond, rude2018policy, stone2016rural} by reducing ``rural'' to ``not urban''.
Poor and inconsistent definitions make rural needs easier to ignore, reflecting the politics of measurement ~\cite{childs2022off, lawrence2024different, pine2015politics}: anything that is not measured is likely to be excluded, forgotten, and rendered invisible.
This is critical given previous research that shows the geographic location of a school---such as region, city, or town---significantly impacts the access and quality of instructional resources ~\cite{panizzon2015impact, abamba2021effects, goode2012beyond, rude2018policy}.
Urban and suburban schools benefit from pooled resources ~\cite{ballou1996condition} and economies of scale ~\cite{azano2015exploring} which make costs and outcomes easier to measure and compare.
Rural schools have structural inequities in school funding tied to geographic isolation and tax bases ~\cite{kolbe2021additional} that increase per-student costs, which standard metrics fail to account for.
The consequences reveal a deeper question about who gets to participate in educational futures, especially when resources are unavailable to invest in advanced technology.
Urban and suburban schools, with their established pipelines to state resources and policymakers~\cite {wood2023review, lawrence2024different}, naturally become early adopters and beneficiaries of GenAI tools.
Yet, this ``natural'' advantage reflects deliberate (or oblivious) processes of ``opportunity hoarding'' and ``digital redlining''~\cite{souto2023hoarding, rury2015opportunity, mccall2022socio}: the systematic concentration of educational innovation within already privileged spaces.
Our findings extend this discussion with one rural educator (\textbf{I6}) sharing that PD opportunities are unlikely to reach them due to their schools' location, leaving educators eager to enhance their teaching without guidance on how to do so effectively.

When rurality remains ill-defined in policy, rural educators---already operating amid systemic inequities---find themselves further pushed to the margins of educational technology innovations and adoption policies~\cite{lawrence2024different}, exemplifying what \textit{rural consciousness} ~\cite{walsh2012putting} identifies as distributive injustice arising from the systematic exclusion of rural concerns by those in power.
Critical decisions are often made \textit{before} the local and affected communities can provide input~\cite{clay2022constructed}, leading to what \textbf{I6} from rural North Carolina described as a ``wait and see'' approach where important and timely decisions on GenAI adoption guidelines and guardrails are deferred to state mandates, limiting local educational communities' ability to proactively address harms and embrace opportunities.

\subsubsection{Beyond GenAI Access: Equity Requires Contextualization}
True equity in educational technology requires more than making tools available;
it requires that innovations be made accessible, sustainable, and adaptable in ways that honor rural values and address structural constraints ~\cite{walsh2012putting}.
Theories like modernization~\cite{nnanna2025role} promote the mindset that technology adoption is a linear path to progress which will automatically improve livelihoods and societal well-being.
Yet, this mindset is problematic for several reasons.

First, GenAI represents a new technology whose educational impacts remain poorly understood~\cite{park2024promise, belghith2024testing, han2024teachers, tan2024more, ibrahim2022understanding}.
Second, while acknowledging its potential benefits, GenAI has raised significant concerns among educators in urban-suburban contexts~\cite{park2024promise, han2024teachers} and rural settings like those in our study.
Many concerns stem from fears that GenAI is indiscriminately revolutionizing traditional practices of educating and curriculum building that rely on human-to-human connection---e.g., discussions, project-based activities, and experiential learning (\textbf{S8, S10, I1}).
This is significant for rural education, where our participants emphasized these pedagogical approaches as central to their practice, raising important concerns about how GenAI could overlook or undervalue these practices.
Finally, GenAI does not always meaningfully complement educators' pedagogical goals rooted in community values.
While some participants (\textbf{S13}) hoped GenAI could help generate ideas for projects that draw on local resources and strengthen location-based learning, central to rural pedagogy, current GenAI systems rarely offer such place-specific features, or educators may be unsure how to achieve this with existing tools.
This disconnect overlooks the sense of belonging from \textit{rural consciousness}, where location-based learning draws on local values and sense of place as educational resources.
Without mechanisms to integrate GenAI into culturally grounded practices, it remains far from rural realities.
Similar to what Cao et al.~\cite{cao2025ai} propose, leveraging cultural capital offers an approach to enrich AI education for underserved communities while aligning with educator's pedagogical goals and supporting the integration of culturally responsive educational technologies.

\subsubsection{Barriers and Buffers from the Digital Divide: Addressing Infrastructural and Systemic Inequities}
Specifically for rural educators, modernization theory's assumption that adoption alone will produce equitable outcomes is also fundamentally flawed due to persistent structural inequities.
Failing to acknowledge and address these existing disparities before and during GenAI implementation risks what \textbf{S17} described as \textit{``exacerbat[ing] the digital divide\footnote{The digital divide refers to the gap between those who have access to and can effectively use digital devices, internet connections, and applications, and those who cannot. This divide also extends to broader information and communication technologies ~\cite{van2020digital}.}''.} 
The risk is reflected in Ladson-Billings' notion of ``educational debt''~\cite{ladson2006achievement}, where historical disinvestment and accumulated neglect towards marginalized groups continue to produce present-day educational disparities.
GenAI's potential in rural areas is constrained by systemic inequities, including inconsistent internet access and limited digital devices, alongside staffing pressures shaped by structural factors like high educator turnover rates and insufficient professional support, as our participants noted.
Moreover, these educators reported fewer opportunities for AI literacy and PD compared to their counterparts in better-resourced contexts.
Until these inequities are addressed, GenAI in rural schools remains limited and the persistent digital divide serves as a potential \textit{loss of opportunity} for effective GenAI integration.

Paradoxically, some educators were observed to experience a sense of relief in their technological isolation (\textbf{I5}).
Several participants noted infrastructure and resource constraints meant GenAI adoption and its effects were not yet an immediate concern, as they were ``protected'' by their own circumstances, which delayed the need to navigate these new challenges.
This dynamic---where engagement with GenAI is shaped by contextual realities---reveals how resources and training in preparation for responding effectively are not fully realized until GenAI becomes relevant.
This delay can also function as a \textit{beneficial buffer}, with opportunities to intentionally prepare and integrate GenAI on rural communities' own terms.
Instead of being immediately thrown into the challenges of GenAI, support can be provided to help educators secure adequate training and develop AI literacy.
Decision-making processes can involve all stakeholders (e.g., parents, community members) to ensure policies and expectations reflect local values.
Schools and districts can use this time to develop strategic policy frameworks and address potential issues, like academic integrity and student overreliance, before GenAI is fully integrated into classrooms.
The buffer also offers rural educators to learn from the successes and failures in early-adopting schools to avoid repeated mistakes and adapt practices that are contextually relevant.
With this approach, a \textit{rural-centered roadmap} can prioritize community values and AI literacy education over a one-size-fits-all, hurried technology adoption.

\subsubsection{The Trade-off Between Design Misalignment and Hype: Rural Educators Reclaiming Agency in GenAI Adoption}
Companies developing GenAI tools further reinforce these power dynamics ~\cite{hearn2024harms, widder2023open} by failing to consider the needs and contexts of educational GenAI stakeholders---a general issue not unique to those from rural communities.
The neglect of end users' differences and needs in design and implementation processes is well-documented in the HCI community, with developers' failure to align with real-world practices, workflows, and constraints often resulting in unintended consequences~\cite{kalluri2020don, do2023s, balayn2023faulty, luo2025reflecting}.
Our results echo previous findings ~\cite{tan2024more, han2024teachers, oh2024exploring}
that GenAI has increased educator workload and responsibilities, creating new demands for educators to distinguish AI-generated from authentic student homework submissions, manage concerns about cheating and plagiarism, and mitigate disruptions to student engagement.
As a result, GenAI technologies have already shifted power dynamics, placing educators in the position of managing risks created by systems they had no role in designing, as opposed to focusing their attention on facilitating education.

On the other hand, our findings show rural educators pursue alternative approaches to GenAI, including selective adoption or outright refusal.
Some educators in our study believed GenAI should be a tool to support them rather than their students (\textbf{S1, S14}), while others dismissed GenAI as irrelevant or ineffective for their teaching needs (\textbf{S5, S16, S29, S24, I6}).
These divergent responses to GenAI challenge the broader AI hype~\cite{bender2026ai}, promoted by tech companies and the media~\cite{taylor2025ai, google2025ai}, that frames GenAI as essential for future success.
Like many technologies before it, GenAI appears to follow Gartner's hype cycle~\cite{linden2003understanding}, now at the \textit{``Peak of Inflated Expectations''} before an inevitable fall into the \textit{``Trough of Disillusionment''}.
This pattern of expected enthusiasm echoes critiques of earlier educational technologies ~\cite{howley1995power, de2023rise, kormos2021rural, digitalEd-equity, assessment-equity} (e.g., web-based technologies like learning management systems) where adoption was treated as imperative and refusal as a failure to prepare students.
These responses also align with literature on technology and AI refusal~\cite{guest2025against, howley1995power, bender2026ai}, which reframes refusal as a deliberate choice rather than an aversion to innovation.
Rural educators' refusal offers a counter-narrative: technology integration should follow critical reflection on pedagogical value, contextual fit, and actual benefits to teaching and learning~\cite{bender2026ai}.
Such scrutiny is essential for reaching the `\textit{`Plateau of Productivity''}, where GenAI can be meaningfully adapted, especially in rural educational contexts.
For rural educators in particular, GenAI adds new burdens in already overtaxed systems, intensifying the demands on educators who are systematically under-supported despite their central role in sustaining local educational communities.
Yet, these same educators (those most skeptical of uncritical adoption) hold the key to meaningful integration.
Centering their voices and concerns is necessary for technologists to develop solutions that serve them over the current hype.


\subsubsection{Shaping the Future: Rural Educators' Agency in Envisioning GenAI Solutions}
Despite constraints imposed by power imbalances and inequity, we found rural educators who choose to adopt are increasingly exercising their agency in how GenAI is integrated in their classrooms, based on their values and sense of place.
Without explicit standards and policies, many educators took the initiative to proactively shape how GenAI is used in their classrooms.
For instance, some educators experimented with GenAI, adapted it to their teaching needs, and developed their own practices of responsible use (\textbf{I3, I4}).

Rural educators actively reimagine how GenAI can support their communities, resisting narratives of \textit{educational abandonment}.
Our study provided a platform for participants to voice their visions of future possibilities of this technology within their educational contexts.
These speculations align with Jasanoff's sociotechnical imaginaries \cite{jasanoff2015future}: visions of how technology could and should shape education.
Our findings reveal future-oriented hopes for AI adaptation addressing rural challenges---GenAI-powered tools to fill instructional gaps from educator shortages and reduce resource disparities.
\autoref{fig:speculations} shows additional examples.
Beyond these applications, GenAI has shown promise to help educators support students with diverse needs, including bilingual learners~\cite{billingsley2024utilizing, zhou2020education} and those with disabilities like students with autism spectrum disorder or learning differences like dyslexia ~\cite{alkan2024role, swargiary2024ai}. 
While survey responses indicated high confidence in educators' abilities to support these students, rural educators' imaginations of GenAI possibilities builds on their existing practices and pedagogical frameworks, serving as a starting point for further innovation.
Speculative design approaches, paired with AI literacy and co-design methods, could unlock more creative GenAI solutions for supporting and enhancing learning opportunities, particularly for students with diverse needs.

\subsection{Recommendations to the HCI Community}
\label{sec:recoms}
Rural schools possess unique strengths and their engagement with researchers is influenced by approaches respecting local values, building trust, and recognizing the central role of their community life.
Given rural communities' critical awareness of the intentions of outsiders or large institutions, successful interventions require partnerships with local organizations and the use of existing social networks for participation ~\cite{hardy2019rural}.
There are existing calls for stronger partnerships between universities and rural communities ~\cite{chambers2024using}.
Following best practices for collaboration, such partnerships should prioritize open communication, clearly defined roles and responsibilities, and the equitable distribution of benefits to mitigate power imbalances and align goals effectively ~\cite{wiggins2025rural, farrell2021practice}.
While these efforts require sustained collaboration across multiple stakeholders and institutions, there are meaningful steps to strengthen GenAI integration in ways responsive to rural contexts.

We see promising pathways forward through two interconnected approaches: (1) comprehensive PD for empowering rural educators with AI literacy and pedagogical strategies to foster students' critical thinking skills, and (2) fundamental shifts toward inclusive decision-making processes to create genuine opportunities for bottom-up discussions and co-design partnerships.
These align with rural educators' expressed need for PD to learn how to use GenAI themselves and how to guide their students on how to use it responsibly (\textbf{S14, S18, S20, S25}).

Participatory approaches\footnote{We also acknowledge the critiques of participatory approaches ~\cite{cleaver1999paradoxes, charitos2025challenges, peters2018participation, birhane2022power}, particularly in classroom settings ~\cite{charitos2025challenges, peters2018participation}.
Including marginalized individuals does not automatically reduce vulnerability, and participation is not inherently beneficial for everyone.
As Cleaver et al. ~\cite{cleaver1999paradoxes} notes, participatory processes can misalign with local norms or reproduce existing power imbalances. 
Peters et al.~\cite{peters2018participation} further critique the tendency for researchers and designers to exert expert authority in participatory workshops, even while facilitating activities meant to be user-driven. 
Instead, they argue for user-led processes, where participants take leadership in shaping outcomes. 
This requires flexibility compared to time-limited formats and offering support empowering participants rather than directing them.} 
provide a strong foundation for this work.
As demonstrated in Ko and Landesman et al.~\cite{ko2025domain}, participatory design in rural contexts should put forth community and shared culture, co-production of knowledge, and trust-building.
These principles foster practical and sustainable solutions to be co-developed with educators and communities.
More recent works employed these approaches to gain insight on local teachers' perspectives in under-resourced areas ~\cite{sun2025live}, support digital inclusion practitioners to enable digital access for digitally excluded populations ~\cite{parnaby2025beyond} and engage rural community members~\cite{ko2025domain} through methods of asset-based design workshops and interviews.
Our study and past work~\cite{hardy2019rural, stone2016rural} also suggest grounding research in educators' own definitions to ensure relevance and contextual understanding.
For example, in our study, participants shared their own definitions of ``rural'' which shaped our understanding of their communities. 
\textbf{Such approaches would represent a meaningful redistribution of power, moving from decisions made \textit{about} rural communities to decisions made \textit{with and by} them.}
\subsection{Limitations and Future Work}
\label{sec:future}
While our study provides valuable insights, its limitations also highlight the ongoing need to center rural communities in research~\cite{sherwood2000has}.
Conducting studies with representative samples from rural communities proved challenging for several reasons. 
First, we had limited access to participants from these regions, requiring us to rely on our networks and fellow educators to spread recruitment information. 
Following best practices ~\cite{hardy2019rural}, we leveraged our existing social networks for participation.
Second, the lack of a unified definition of ``rural'' created confusion about study eligibility when we advertised to potential participants.
These access challenges limited our study's scope in multiple ways, resulting in a smaller sample size than we had hoped for, and amplifying voices of those who were already in-network and had access to our recruitment materials.
While our results highlight significant challenges with technology and internet access in rural schools, our study required reliable technology to participate, which may have unintentionally limited the inclusion of educators most directly navigating connectivity challenges.
This underscores the importance of designing research methods that consider these realities.
Due to reliance on in-network recruitment, our rural geographic scope covered three states, which may have restricted the generalization of our findings.
Despite all these limitations, the number and quality of responses in our study provide a meaningful contribution to share voices from the rural educators.
Future work should expand the sample size and include additional states and rural regions to ensure broader representation of rural educators.
Additionally, although the survey was open to all disciplines, the high response rate from STEM educators could suggest a self-selection bias ~\cite{kazmierczak2023self}, as they may be more familiar with GenAI and therefore more inclined to participate.
Expanding recruitment strategies to research only non-STEM educators can offer different view-points from educators who may use technology less in their everyday practice due to the subject matter, though it might not be as easy or possible, as many educators within rural communities teach multiple subjects ~\cite{rude2018policy, zinger2020teaching, barley2009preparing, jones2025integrating}. 
While our study focuses specifically on the perspectives of educators, a fuller understanding of GenAI's role in education can be explored with the voices of students, parents, and community members, since their values and practices also shape what meaningful adoption looks like.

Supported by our findings, there is currently a significant gap and opportunity for researchers to collaborate with rural educators on ways to expand meaningful GenAI use, such as efforts to develop AI literacy curricula and professional development opportunities, specifically focused on responsible and safe adoption for both educators and students.
Such efforts can equip rural educators with a meaningful skill set to participate in technological practices and challenge deficit-based framings that have historically excluded them from conversations about emerging technologies driving educational innovation. 

Our interview participants proposed ideas for GenAI tools tailored to their specific needs. 
We hope researchers and technologists will be inspired by these findings to focus on prototyping and co-designing rural-focused tools with rural education stakeholders, ensuring GenAI-powered educational tools are practical, effective, and capable of enhancing teaching and learning for those within these communities.
Potential approaches could include locally-hosted language and diffusion models that address connectivity concerns and collaborative GenAI interfaces, building educator confidence while preventing overreliance. 
However, the most effective solutions will emerge from genuine collaboration and co-design with rural stakeholders themselves.
We emphasize \textbf{designing tools with rural stakeholders' needs in mind does not limit their broader applicability. 
Rather, supporting rural educators could benefit other educators, whether facing similar resource constraints---including those in low-income urban and suburban schools ~\cite{cao2025ai, thomas2024improving}---or trying to navigate the new reality AI has created.} 
This inclusive design approach has the potential to create more equitable educational technologies that serve diverse learning environments.



\begin{acks}
We thank Lucy Havens for her invaluable support and insightful feedback on this manuscript, especially for generously sharing her expertise in critical theory to strengthen our analysis and discussion. We also thank our collective network who helped us with recruitment; especially, we thank Mike Vargas from Arizona, Alicia Jalbert from the Roux Institute at Northeastern University, Ian Collins from Maine Mathematics and Science Alliance (MMSA), and Amy Rice and Janelle Ehnstrom from North Carolina for their invaluable assistance in connecting us with participants from rural communities.
Most importantly, we extend our deepest gratitude to the rural educators who generously shared their time and insights despite their demanding schedules to make this study possible.
Finally, we thank the members of Northeastern University's Khoury Visualization Lab and The Roux Institute's Human-Data Interaction Group for their feedback on this work.

This work was supported by the National Science Foundation (NSF) through grant DRL-2318339. 
Any opinions, findings, and conclusions or recommendations expressed in this material are those of the authors and do not
necessarily reflect the views, opinions, or policy of the NSF.
\end{acks}
\bibliographystyle{ACM-Reference-Format}
\bibliography{Bibliography}



\end{document}